# Matrix models as solvable glass models

L. F. Cugliandolo *, J. Kurchan †, G. Parisi

*Dipartimento di Fisica, Università di Roma I, La Sapienza, Roma, Italy,*

*INFN - Sezione di Roma I*

F. Ritort

*Dipartimento di Fisica, Università di Roma II, Tor Vergata, Roma, Italy,*

*INFN - Sezione di Roma I*

(July 21, 1994)

## Abstract

We present a family of solvable models of interacting particles in high dimensionalities without quenched disorder. We show that the models have a glassy regime with aging effects. The interaction is controlled by a parameter $p$. For $p = 2$ we obtain matrix models and for $p > 2$ 'tensor' models. We concentrate on the cases $p = 2$ which we study analytically and numerically.

75.10.Nr, 64.60.Cn, 64.70.Pf, 11.17.+y

Typeset using REVTEX

---

*Address after October 1st: Service de Physique de l'Etat Condensé, CEA, Saclay, France.

†Address after October 1st: LPTHE, Ecole Normale Supérieure, Paris, France.

Disordered systems such as structurally disordered materials (glasses) and spin-glasses are characterized by complex free-energy landscapes and non-equilibrium phenomena with time-scales exceeding the experimental times [1–5]. The glass transition being mainly kinetic, one has to face the full dynamical problem in order to grasp many aspects of the physical behavior of these systems. This is a very difficult problem for realistic models.

As regards to *spin*-glass systems, it has been recently noticed that simple yet microscopically inspired models, for which the mean-field solution is exact [4], have both solvable and non-trivial dynamics which agrees at least qualitatively with that of realistic systems [6–8]. Spin-glasses are described by Hamiltonians with quenched (random) disorder included *by hand* and do not have an ordered ground state to which the system can possibly decay. Structural glasses instead have an ordered ground-state though when cooled sufficiently fast they remain in a regime of (self-induced) disorder. Various aspects of the glassy behavior have been studied with phenomenological scenarios, dynamical theories, toy kinetic models and numerical simulations (see *e.g.* refs. [1–3]).

However, unlike the case of *spin*-glasses, there has been a lack of simple, solvable Hamiltonians capturing the essentials of structural glasses. Several recent papers investigate mean-field models which despite having no explicit quenched disorder still have a glassy transition [9,10,12]. In particular, the model of Ref. [12] has been shown to obey dynamical equations identical to those of the $p$-spin glass solved in Ref. [6]. It is not clear whether the emergence of such glassy behaviour in these models is a consequence of using 'complicated' (*i.e.* quasi random) Hamiltonians. The object of this note is to present a family of very simple models that can be interpreted as interacting particles in high dimensionalities and are suitable to describe the supercooled liquid phase and the glassy regime.

We consider a system of $n$ particles moving in an $N$-dimensional space, in the limit $n, N \to \infty$ [13]. In terms of the coordinates $s^a$ of the $a$-th particle the Hamiltonian reads:

$$H \equiv \frac{1}{n} \sum_{a \neq b} (s^a \cdot s^b)^p ,\qquad(1)$$

where $s^a = (s_1^a, ..., s_N^a)$ and $a, b = 1, ..., n$. We consider the case ($\pm$) in which the particles are constrained to move on a spherical surface $|s^a|^2 = N \; \forall a$, and the case in which they can occupy only the vertices of a hypercube $s_i^a = \pm 1$. The constraints counterbalance the repulsive force and confine the particles. The range of the interaction is smaller the larger the value of $p$. We choose $n$ such that the force term $\beta \nabla^a E$ on each particle $a$ has components of order one and involves all other particles if they are distributed randomly (i.e. if $(s^a \cdot s^b) = O(\sqrt{N})$). This implies the set of choices $\beta \sim O(\sqrt{\frac{n}{N^{p-1}}})$ and, in particular, we use $n = 2\alpha N^{p-1}/p!$ with $\beta \sim O(1)$ or $n = 2\alpha N/p!$ with $\beta \sim O(N^{1-p/2})$, both of which yield non-trivial models.

In this paper we concentrate in the model with $p = 2$:

$$Z = e^{\beta N^2} \text{Tr}_{\{s\}} \exp\left[-\frac{\beta}{\alpha N} \text{tr}(S^\dagger S S^\dagger S)\right] \qquad (2)$$

where $S$ is the $N \times n$ rectangular matrix of elements $s_i^a$, and $\text{Tr}_{\{s\}}$ runs over either the spherical or the Ising measure. The cases $\alpha > 1$ and $\alpha < 1$ are related; if $\alpha > 1$ the free-energy satisfies $F(\alpha, \beta) = 1/\alpha \; F(1/\alpha, \tilde{\beta} = \alpha\beta) + (\alpha - 1)N^2$. We also report very briefly some results for the model with $p = 3, 4$ and, furthermore, for the $\pm$ version of (2) but with *symmetrical S*.

The partition function (2) is that of a rectangular matrix model. In the spherical case it can be solved exactly; it does not have a glassy phase transition but behaves like a 'liquid' at all temperatures. The $\pm$ model has, instead, three phases:

*i.* A high-temperature ('liquid') phase in which its free-energy coincides (up to a constant) with that of the spherical model.

*ii.* Low-temperature $T < T_f$ ('crystalline') states that dominate the low-temperature Gibbs-measure. The statical transition is first order.

*iii.* A dynamical low temperature ('glass') regime $T < T_g$.

Our strategy is as follows: we first show the equivalence between spherical and $\pm$ models in the high-temperature phase, and we solve the spherical model for all temperatures. Then we briefly discuss the nature of the 'crystalline' states of the $\pm$ model. Finally, we verify numerically, using Montecarlo dynamics, these analytical results. From the simulations it becomes clear that the ± model has a dynamical transition at a temperature $T_g$: below $T_g$ a large enough system asymptotically reaches an energy that is higher than the analytical continuation of the high-temperature phase, and it never (in the limit $N \to \infty$) 'falls' into a 'crystalline' state. Furthermore, we show that the dynamics in the 'glass' phase is remarkably similar to that of both mean-field and realistic *spin*-glasses. We expect that similar results should be obtained with *e.g.* Langevin or Glauber dynamics.

Let us first perform a Gaussian decoupling of the quadratic term in Eq. (2) by introducing auxiliary variables $M_{ab}$:

$$Z = \chi(\beta) \int DM \; \text{Tr}_{\{s^a\}} \exp\left(-i\sqrt{\frac{\beta}{\alpha}} \sum_{a,b} M_{ab} s^a \cdot s^b\right) . \qquad (3)$$

The measure $DM$ depends on the $M$ ensemble we consider, both choices $DM \equiv \prod_{a<b} dM_{ab} \sqrt{N/4\pi} \exp(-N/4 \; \text{tr} M^2)$ ($M$ real, symmetric with zero diagonal) and $DM \equiv \prod_{a \leq b} dRe(M_{ab}) \sqrt{N/4\pi} \prod_{a<b} dIm(M_{ab}) \sqrt{N/4\pi} \exp(-N/4 \; \text{tr} M^2)$ ($M$ hermitian) are possible. Then $\chi = 1$ and $\chi = e^{\beta N^2}$, respectively.

For the spherical model we have

$$Z_{sph} = \chi(\beta) \int \prod_a \frac{dz^a}{\sqrt{2\pi}} \exp\left[\frac{iN\beta}{2\alpha} \sum_a z^a\right] \int DM \exp\left[-\frac{N}{2} \text{tr} \ln\left(i\frac{\beta}{\alpha}\delta_{ab} z^a + 2i\sqrt{\frac{\beta}{\alpha}} M_{ab}\right)\right] . \qquad (4)$$

Since $z^a$ are only $N$ variables and the exponent is symmetrical w.r.t. the permutations of the indices of $z^a$, an admissible saddle point is $-iz^a \mid_{s.p.} = \mu, \; \forall a$. This implies that the solution coincides, in the large $N$ limit, with that of the same model with the *global* constraint: $\sum_{i,a} (s_i^a)^2 = nN$.

The model written in this form, with hermitian $M$, can be recognized as a form of the generalized Penner model [14] or of the Penner-Kontsevich model before integrating over $z^a$ [15].

We now prove that the high-temperature expansions for the free-energies of the spherical and the $\pm$ models coincide to all orders in $\beta$. Let us go back to the $\pm$ version of Eq. (3), and add a term $0 = \frac{1}{2}\sum_a \nu_a(N - |s^a|^2)$. We later fix $\nu_a$ appropriately. With the symmetric decoupling Eq. (3) becomes:

$$Z_\pm = e^{\frac{N}{2}\sum_a \nu_a} \int DM \, \exp\left[N \ln \text{Tr}_{\{s^a\}} \exp\left(-\frac{1}{2}\sum_{ab} s^a D_{ab} s^b\right)\right], \qquad (5)$$

with $D_{ab} = 2i\sqrt{\beta/\alpha}M_{ab} + \delta_{ab}\nu_a$. Performing the trace over the $s^a$ after a Gaussian decoupling the exponent in (5), $NW(M,\nu)$, is

$$W(M,\nu) = -\frac{1}{2}\text{tr}\ln(D+I) + n\ln 2 + \ln\int \prod_a \lambda_a \exp\left(-\frac{1}{2}\sum_{ab}\lambda_a \Delta^{-1}{}_{ab}\lambda_b - \sum_a \frac{\lambda_a^4}{12} + \ldots\right) \qquad (6)$$

where $\Delta^{-1} \equiv D^{-1} + I$ and the dots stand for the higher orders of the series of $\ln(\cos\lambda_a)$.

The off-diagonal elements of the propagator $\Delta_{ab}$ are $O(1/\sqrt{N})$. We now choose the $\nu_a$ to impose that the diagonal elements are also of that order. This yields for $\nu_a$ the implicit relations $[D+I]^{-1}_{aa} = 1$, and $W(M,\nu(M))$ becomes

$$W(M,\nu) = \frac{N}{2}\sum_a \nu_a - \frac{1}{2}\text{tr}\ln(D+I) + n\ln 2 + \text{`diag'}(M) \qquad (7)$$

The term 'diag', stands for the diagrams generated by the higher orders in $\lambda_a$ in (6) with the propagator $\Delta$. These are functions of $M$ which can then be treated as perturbations of $\text{tr}\,M^2$ in the integral over $M$ (for example, the first term is $\sum_{ab}[M_{ab}]^4$). They are in turn vertices of the 'fat diagrams' (with propagator $\text{tr}\,M^2$) which lead to non-planar diagrams, and hence can be neglected in the large $N$ limit (order by order in $\beta$). This is in contradistinction to the vertices in the 'fat diagrams' coming from the trace of the logarithm in (7) (of the form $\text{tr}\,M^k$) which cannot be neglected for any $k$.

Identifying $\nu_a \leftrightarrow i\beta z_a/\alpha - 1$, the condition for the $\nu^a$ now coincides with the saddle point equation for the $z^a$ in the spherical model. Hence both free energies coincide to leading order in $N$, up to an additive constant in the entropy.

In order to solve the spherical model we return to Eq. (2). Assuming $\alpha \geq 1$ and using the large $N$ equivalence with the global constraint we get, up to constant multiplicative factors:

$$Z_{sph} \sim e^{\beta N^2} \int d\mu \int \prod_{i=1}^{N} dx_i \, \exp\left(-N^2 \frac{\beta\mu}{2} - \beta E[x_i]\right). \tag{8}$$

The $x_i\sqrt{N}$ are the 'diagonal' values of $S$ in its canonical form and $E[x_i]$ reads

$$E[x] = N \sum_{i=1}^{N} \left(\frac{1}{\alpha}x_i^4 - \frac{\mu}{2\alpha}x_i^2 - \frac{(\alpha-1)}{\beta} \ln |x_i|\right) - \frac{1}{2\beta} \sum_{i \neq j} \ln |x_i^2 - x_j^2|. \tag{9}$$

We define a symmetric density of eigenvalues $\rho(x) = \frac{1}{2}\sum_i [\delta(x-x_i) + \delta(x+x_i)]$ to get the equation (in the continuous limit)

$$\frac{4x^3}{g} - \frac{\mu x}{g} - (\alpha - 1) = \int dy \frac{\rho(y)}{x-y}, \tag{10}$$

where $g \equiv 2\alpha T$. Following Refs. [16,17] we propose the generating function

$$F(z) = \frac{4z^3}{g} - \frac{\mu z}{g} - \frac{(\alpha-1)}{z} - \frac{4z}{g}\sqrt{(z^2-a^2)(z^2-b^2)}(z^2 + c^2). \tag{11}$$

The jump across the cut of $F$ yields $\rho$. The constants $a$, $b$, $c$ and the saddle point value of $\mu$ are determined by the conditions: *i.* $\rho \sim 1/z$ as $z \to \infty$ (normalization), *ii.* $F(z)$ has no pole in $z = 0$ and *iii.* the constraint $\sum x_i^2 = n = \alpha N$. Defining $a_{\pm} \equiv a^2 \pm b^2$, these conditions yield fifth degree polynomials for $a_+$, $a_-^2$, $c^2$ and $\mu$. The energy density is obtained from a contour integral of $z^4 F(z)$ around the cut and gives $\epsilon \equiv E/(nN) = (1/\alpha)[-1 + a_-^2/(32g\alpha)(a_-^2 + 4a_+^2 + 8a_+c^2)]$. In fig. 1 we plot $\epsilon$ vs. $T$ for $\alpha = 1$.

At $T \sim 0$, $E = N^2[(\alpha-1) + T/4]$ for $\alpha > 1$ (and $E = N^2 T/(4\alpha^2)$ for $\alpha < 1$), showing explicitly that the entropy tends to $-\infty$ logarithmically at $T = 0$. Then the entropy of the high-temperature solution vanishes at some temperature $T_o > 0$ (for $\alpha = 1$, $T_o \sim 0.15$) and there must be a temperature $T_f \geq T_o$ in which the $\pm$ system undergoes a *static* transition. Two possibilities then arise: either the low-$T$ partition function is dominated by a glassy state or by an ordered 'crystalline state'. For the spherical model a little algebra shows that the ground states are obtained by constructing the $N \times n$ matrix $S$ with rows (columns) of orthogonal vectors of norm $\sqrt{N}$ ($\sqrt{n}$) for $n > N$ ($N > n$). For the $\pm$ model, we are not always able to construct such a matrix with $\pm 1$ elements, though for some values of $n, N$ this is possible (e.g. $n = N = 2^k$), in which case the energy of the lowest configuration coincides with that of the spherical model. For these $N, n$ these states dominate the low-$T$ partition function; for other values we have observed numerically other states which though non-optimal play the same role at low-$T$.

Since these states have very small entropy, $T_f$ is approximately the temperature at which the free-energy of the high-temperature coincides with their energy (for $\alpha = 1$, $T_f \sim 0.41$).

The important point is that these 'crystalline' ground states are different from the glassy regime and are not reached dynamically by large enough systems, as we shall see.

In fig. 1 we show the analytic and numerical results for the temperature dependence of the energy. The numerical points were obtained by means of a slow annealing in temperature, the energy densities obtained in this way ('dynamical energies') tend to have a well-defined limit with $N$, for $N$ large enough. For the spherical model the 'dynamical energy' and the analytically predicted static energy coincide for all temperatures. In the high temperature phase the 'dynamical energies' of the spherical and $\pm 1$ models also coincide.

Below the glass temperature $T_g \sim .5$ ($> T_f$), the 'dynamical energy' of the $\pm$ model is higher than that of the spherical model. For the cases in which the ground state is known (e.g. $N = n = 2^k$) we have checked that the system takes a time that grows fast with $N$ to 'fall' into a crystalline state.

In order to undertand the nature of the 'glass' phase, we have studied the decay of the auto-correlation functions $C(t, t') \equiv 1/(Nn) \langle \sum_a s^a(t) \cdot s^a(t') \rangle$. Above $T_g$ for the $\pm$ model, and at all temperatures for the spherical model, the averaged auto-correlations decay with a time translationally-invariant law ($C(t + t_w, t_w) = C(t)$) as in a system in equilibrium. Below $T_g$ the $\pm$ model presents aging effects [5]: the auto-correlation functions shown in fig. 2 are clearly those of a system away from equilibrium even for very long times. The auto-correlations $C(t + t_w, t_w)$ have, for large $t, t_w$, an approximate $t/t_w$ dependence (for large $t/t_w$ $C \sim (t/t_w)^{-.05}$) and are strikingly similar to the corresponding ones of mean-field [6–8] and $d = 3$ *spin*-glasses [18].

Let us mention in passing that we have performed dynamical simulations of the $\pm$ model (1) with $p = 3$ ($n \sim N^2, \beta \sim O(1)$), with $p = 4$ ($n \sim N, \beta \sim O(1/N)$) and also of (2) with $S^\dagger = S$. In all these cases we have found glassy behavior.

In order to make contact with the theory of systems with quenched disorder we first note that the if in the system described by the Eq. (1) we study the dynamics of one particle, while fixing the other $n - 1$ particles in random positions, we obtain neural network models (at negative temperature) [19]; systems known to have a *spin*-glass behaviour.

A different strategy, that can be used to solve all the models we have described, consists in proposing a model with quenched disorder which is expected to have the same dynamical behavior [9,10]. For the case $p = 2$ the natural choice is to consider the Hamiltonian (1) with the $s_i^a$ defined by $s_i^a = U(ai, bj) \, t_j^b$, where $t_i^a$ are either spherically constrained or $\pm 1$. The tensor $U(ai, bj)$ is a quenched random variable taken from the ensemble of orthogonal transformations $\mathcal{R}^{n \times N} \to \mathcal{R}^{n \times N}$ (*i.e.* preserving the norm $\| t \|^2 \equiv \sum (t_i^a)^2$). One then considers (quenched) averaged quantities over $U$.

The analysis of the statics uses the replica trick and goes along the lines of Ref. [10]. We do not describe it in detail here, but report some results: If the $t_i^a$ are spherically constrained the disordered model coincides with the non-disordered spherical model for all temperatures. If the $t_i^b$ take values $\pm 1$, the high-temperature phase (uncoupled replicas) also coincides with the spherical model. In order to analize the low-temperature (glassy) phase of the disordered model one ends up with a replicated matrix model.

Perhaps more ambitiously, one could try to face the full dynamical problem as developed for mean-field spin glasses [6].

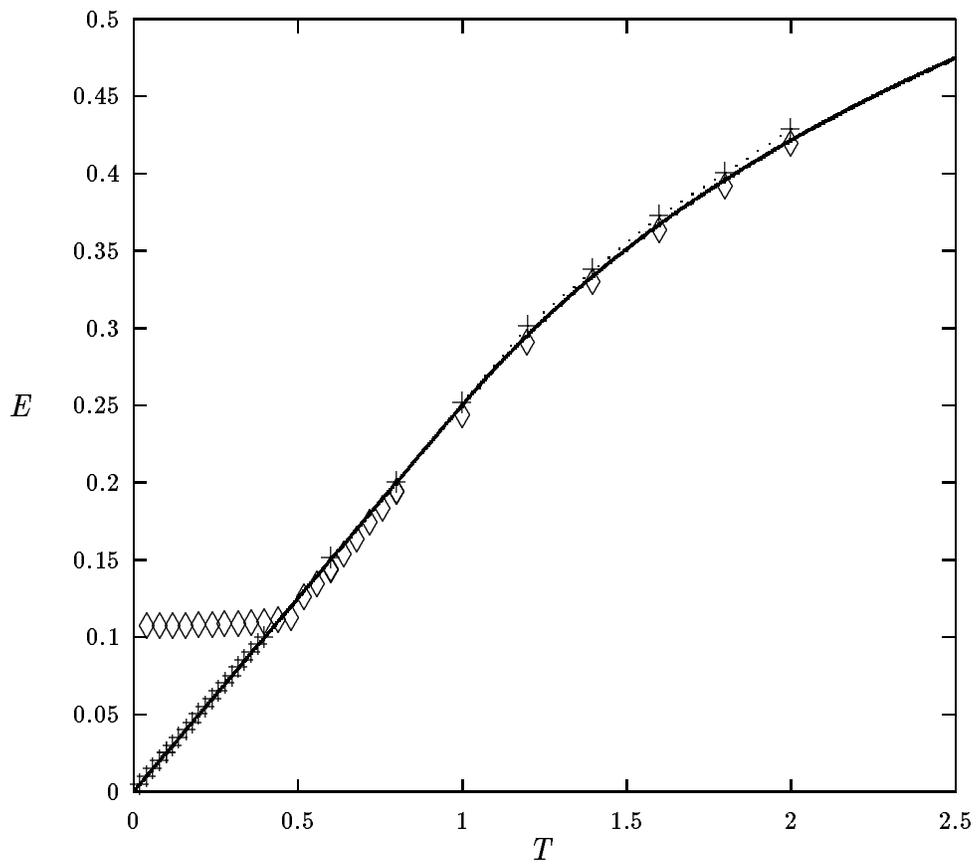

**Fig. 1** Numerical values for the energy density *vs.* temperature for the spherical (crosses) and ± (diamonds) models. $N = n = 50$, $\alpha = 1$. The solid line shows the analytical solution for the spherical model.

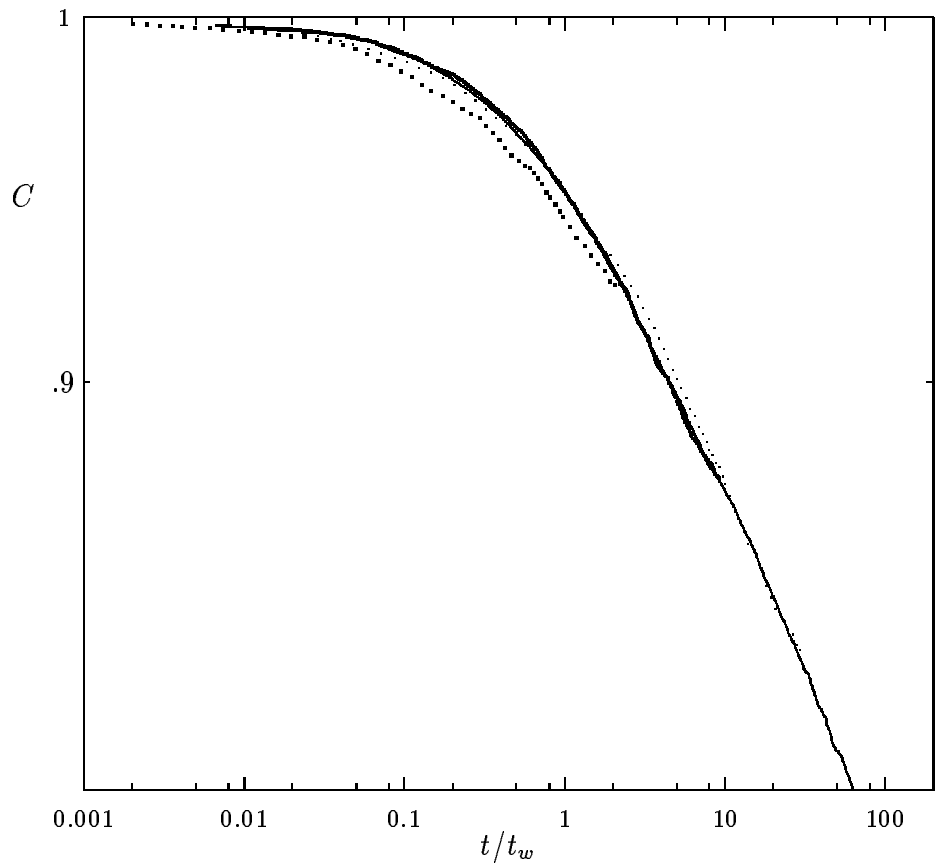

**Fig. 2** Aging curves for the ± model for $\alpha = 1$ at $T = 0.2$ ($< T_g$); $\ln - \ln$ plot, $C(t + t_w, t_w)$ vs. $(t/t_w)$ for $t_w = 1000, 300, 100, 30$, $t$ up to 3000. $N = n = 500$.

We wish to acknowledge useful suggestions from S. Franz, E. Marinari, J. Ruiz-Lorenzo and G. Zemba.